\documentclass{iopart}
\usepackage{graphicx}
\usepackage{epsfig}
\pdfoutput=1
\usepackage{iopams}
\usepackage{setstack}

\usepackage{mathrsfs}
\usepackage{amsfonts}
\usepackage{amssymb}
\begin{document}

\title[The A-Cycle Problem for Transverse Ising Ring]
{The A-Cycle Problem for Transverse Ising Ring}

\author{Jian-Jun Dong$^1$$^,$$^2$, Peng Li$^1$$^,$$^2$ and Qi-Hui Chen$^3$}

\address{$^1$ College of Physical Science and Technology, Sichuan University, 610064,
Chengdu, China}
\address{$^2$ Key Laboratory of High Energy Density Physics and Technology of Ministry of
Education, Sichuan University, Chengdu, 610064, China}
\address{$^3$ Key Laboratory of Advanced Technologies of Materials (Ministry of Education of
China). Superconductivity R$\&$D Center (SRDC), Mail Stop 165$\#$, Southwest
Jiaotong University, Chengdu, Sichuan 610031, China}

\ead{lipeng@scu.edu.cn}

%\date{\today}

\begin{abstract}
Traditionally, the transverse Ising model is mapped to the fermionic c-cycle
problem, which neglects the boundary effect due to thermodynamic limit. If
persisting on a perfect periodic boundary condition, we can get a so-called a-cycle
problem that has not been treated seriously so far (Lieb \textit{et al.}, 1961 \textit{Ann. of Phys.}
\textbf{16} 407).
In this work, we show a little surprising but exact result in this respect. We find
the odevity of the number of lattice sites, $N$, in the a-cycle problem plays an
unexpected role even in the thermodynamic limit, $N\rightarrow\infty$, due to the
boundary constraint. We pay a special attention to the system with
$N(\in Odd)\rightarrow\infty$, which is in contrast to the one with
$N(\in Even)\rightarrow\infty$, because the former suffers a ring frustration.
As a new effect, we find the ring frustration induces a low-energy gapless
spectrum above the ground state. By proving a theorem for a new type of Toeplitz
determinant, we demonstrate that the ground state in the gapless region exhibits
a peculiar longitudinal spin-spin correlation. The entangled nature of
the ground state is also disclosed by the evaluation of its entanglement
entropy. At low temperature, new behavior of specific heat is predicted.
We also propose an experimental protocol for observing the new phenomenon
due to the ring frustration.
\end{abstract}

\maketitle

%\vskip 2cm

\section{Introduction}

Ising spin chain in a transverse field%
\begin{equation}
H=J\sum_{j=1}^{N}\sigma_{j}^{x}\sigma_{j+1}^{x}-h\sum_{j=1}^{N}\sigma_{j}^{z}
\label{H1}%
\end{equation}
with Pauli matrices $\sigma_{j}^{\alpha}$ ($\alpha=x,z$) is a well-known
prototype for demonstrating quantum phase transition \cite{Sachdev}.
Jordan-Wigner transformation~can be employed to solve it
\cite{J-W,Lieb,Pfeuty}. By neglecting the boundary effect in thermodynamic
limit, Lieb \textit{et al}. defined and solved a "c-cycle" problem. While the
original problem without any approximation is called an "a-cycle" one
\cite{Lieb,Suzuki}. In the c-cycle problem, the thermodynamic limit is
performed at the beginning, which brings the model to a free fermion problem.
While in the a-cycle problem, if we consider a perfect periodic boundary
condition (PBC) for the original spin model, we get a constraint fermion
problem and have to keep an arbitray $N$ during the calculation. We only have
the opportunity to take the limit $N\rightarrow\infty$ at the end of calculation.

The theoretical properties of the model (\ref{H1}) have been related to real
materials since decades ago \cite{Suzuki,Dutta}. Researchers have also been
looking for nowadays state-of-art techniques, such as the ones based on
laser-cooled and trapped atomic ions, to mimic this model \cite{Edwards,Kim}.
But these artificial systems can only produce finite lattices in principle,
through which we hope to see the trend for large enough systems. The system
with perfect PBC can be realized through a ring geometric optical lattices
\cite{Amico}, which demands a thorough comprehension of the a-cycle problem.
To the best of our knowledge, it has not been treated seriously up to now
\cite{Lieb}. In this work, we shall develop a systematic method of band
structure analysis to handle it and produce exact result that can match the
full degrees of freedom of the spin model. The results will also be confirmed
by an alternative method of exact diagonalization on small systems.

%%%%%%%%%%% FIG 1  %%%%%%%%%%%%%%%%%%%%%%%%%%%%%%%

\begin{figure}[h]
\begin{center}
\includegraphics[width=3.2in,angle=0]{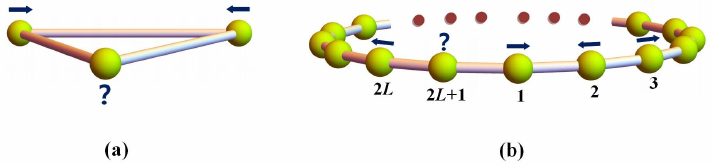}
\end{center}
\caption{The main focus of this paper: transverse Ising ring with odd number
of lattice sites: (a) $N=3$, (b) $N=2L+1$ ($L=1,2,3,\cdots$). The Hamiltonian
is shown in (\ref{H1}), which exhibits a ring frustration due to
antiferromagnetic seam ($J>0$).}%
\label{fig1}%
\end{figure}

%%%%%%%%%%%%%%%%%%%%%%%%%%%%%%%%%%%%%%%%%%%%%%

On the other hand, frustration is an intriguing topic. Very few frustrated
models can be solved exactly \cite{Diep}. To seek for nontrivial phenomenon,
we mainly focus on the antiferromagnetic ($J>0$) system with PBC and large
enough $N\in Odd$, because it suffers a \textit{ring frustration}
\cite{Solomon,Owerre} as a result of antiferromagnetic seam \cite{McCoy Wu}
(please see Fig. \ref{fig1}). Notice the ring frustration here is not a short
range type of the usual case. One may ask whether the odevity of $N$ plays a
meaningful character in the a-cycle problem when $N\rightarrow\infty$. The
answer is affirmative. By the rigorous solution, we demonstrate that the
combination of the a-cycle problem and ring frustration does result in a
dramatical consequence. To understand the fascinating result in a contrastive
manner, we also discuss the system without any frustration, i.e. with $N\in
Even$.

We arrange the contents of this paper as follows. In Section \ref{sec:JW}, we
discuss the details about how the periodic spin problem is turned into a
fermionic a-cycle one. In Section \ref{sec:even}, we dwell on the a-cycle
problem without ring frustration ($N\in Even$). We develop the method of band
structure analysis. In Section \ref{sec:odd}, we work on the a-cycle problem
with ring frustration ($N\in Odd$). We demonstrate that the presence of ring
frustration will induce an interesting gapless spectrum above the ground state
in the strong antiferromagnetic region. We
demonstrate that the ground state exhibits a strong longitudinal spin-spin
correlation and possesses a considerably large entropy of entanglement.
We also give finite-temperature properties of the gapless region,
including the density of states (DOS) and the specific heat. In Section
\ref{sec:exp}, we propose an experimental protocol with special concern of the
realization of ring frustration. At last, we give a discussion in Section
\ref{sec:conclusion}.

\section{Jordan-Wigner fermions and the statement of the a-cycle problem}

\label{sec:JW}

It is convenient to convert the Pauli matrices to the raising and lowering
operators,
\begin{equation}
\sigma_{j}^{x}=\sigma_{j}^{+}+\sigma_{j}^{-},\sigma_{j}^{z}=2\sigma_{j}%
^{+}\sigma_{j}^{-}-1.
\end{equation}
By introducing the Jordan-Wigner fermions that abide by the non-local
relations,%
\begin{equation}
\sigma_{1}^{+}=c_{1}^{\dag},\quad\sigma_{j}^{+}=c_{j}^{\dag}\mathrm{exp}%
(\mathrm{i}\pi\sum_{l<j}c_{l}^{\dag}c_{l}),
\end{equation}
the spin model, (\ref{H1}), can be transformed to
\begin{eqnarray}
 H = Nh-2h\sum_{j=1}^{N}c_{j}^{\dag}c_{j}+J\sum_{j=1}^{N-1}(c_{j}^{\dag
}-c_{j})(c_{j+1}^{\dag}+c_{j+1})\nonumber\\
\quad-J\exp(\mathrm{i}\pi M)(c_{N}^{\dag}-c_{N})(c_{1}^{\dag}+c_{1}),
\label{H2}
\end{eqnarray}
where the total number of fermions,
$
M=\sum_{j=1}^{N}c_{j}^{\dag}c_{j},
$
does not conserve. But the parity of the system does, which is defined as
\begin{equation}
P=\exp(\mathrm{i}\pi M)=(-1)^{M}. \label{P}%
\end{equation}
The vacuum state, devoid of any fermions, corresponds to the full polarized
spin state (spin down),%
\begin{equation}
\left\vert 0\right\rangle =\left\vert \downarrow\downarrow\downarrow
\cdots\downarrow\right\rangle . \label{vacuum}%
\end{equation}
(\ref{H2}) defines the full a-cycle problem \cite{Lieb,Dutta}.

As a comparison, the c-cycle problem is defined by neglecting the last term,
$-J[\exp($i$\pi M)+1](c_{N}^{\dag}-c_{N})(c_{1}^{\dag}+c_{1})$, in \cite{Lieb}%
\begin{eqnarray}
H = Nh-2h\sum_{j=1}^{N}c_{j}^{\dag}c_{j}+J\sum_{j=1}^{N}(c_{j}^{\dag}%
-c_{j})(c_{j+1}^{\dag}+c_{j+1})\nonumber\\
\quad-J[\exp(\mathrm{i}\pi M)+1](c_{N}^{\dag}-c_{N})(c_{1}^{\dag}+c_{1}).
\label{H-c-cycle}%
\end{eqnarray}
In doing so, one has accomplished the thermodynamic limit. Thus the c-cycle
problem becomes a free fermion one \cite{Pfeuty}.

While for a system with perfect ring geometry, there holds a precise condition
on the spins, $\sigma_{N+j}^{\alpha}=\sigma_{j}^{\alpha}$. With no ends (or
boundaries) existing, nothing in (\ref{H2}) could be neglected. We have to
keep an arbitrary $N$ in the calculation. We hope to get a result containing
$N$\quad as a variable thus it facilitates us to take the limit,
$N\rightarrow\infty$. Then, we can discern the different consequences of the
limits, $N$($\in Even$)$\rightarrow\infty$ and $N$($\in Odd$)$\rightarrow
\infty$.

One should notice that, although there holds a PBC for the spin operators, a
priori PBC should not be imposed on the fermions since an anti-PBC is also a
reasonable choice. We will demonstrate both of them, $c_{N+j}=c_{j}$ (PBC) and
$c_{N+j}=-c_{j}$ (anti-PBC), are indispensable to restore the full degrees of
freedom of the original spin model, (\ref{H1}), exactly. We will see that the
parity, $P$, will be fixed by the boundary condition of the fermions.

\section{A-cycle problem without ring frustration ($N\in Even$)}

\label{sec:even}

Let us see the case without ring frustration first. To make the Fourier
transformation
\begin{equation}
c_{q}=\frac{1}{\sqrt{N}}\sum_{j=1}^{N}c_{j}\exp(\mathrm{i\thinspace}q\,j)
\end{equation}
available for solving the fermionic problem, we found the boundary condition
must be bound up with the parity. So there are two routes to be followed. When
$M\in odd$, we call it the \emph{odd channel (o)} and when $M\in even$, the
\emph{even channel (e)} respectively. The procedure inevitably becomes a
little tedious. In the following, we delicately use notations to make the
deductions as clear as possible. For example, we use the notations $N\in Even
$ and $M\in even$, although $Even$ and $even$ are the same thing.

In fact, for the case of $N\in Even$, Schultz \textit{et al}. \cite{Schultz}
had discussed the contribution of the two channels in context of the classical
two-dimensional Ising model in the same essence. But their discussion on the
thermal states of the classical two-dimensional Ising model only corresponds
to the ground state property of the quantum transverse Ising model. In this
section, we discuss the quantum model directly and develop a method of band
structure analysis for both ground states and excitations.

\subsection{Diagonalization in the odd channel}

In the odd channel ($M\in odd$), the Jordan-Wigner fermions must obey PBC:
$c_{N+j}=c_{j}$, and the momentum in the first Brillouin zone ($1$st BZ) must
take a value in the set
\begin{equation}
q^{(E,o)}=\{-\frac{N-2}{N}\pi,\ldots,-\frac{2}{N}\pi,0,\frac{2}{N}\pi
,\ldots,\frac{N-2}{N}\pi,\pi\},
\end{equation}
where the superscript $(E,o)$ denotes $N\in Even$ $(E)$ and $M\in odd$ $(o)$.
After the Fourier transformation, the Hamiltonian can be diagonalized by the
Bogoliubov transformation
\begin{equation}
\eta_{q}=u_{q}c_{q}-\mathrm{i}v_{q}c_{-q}^{\dagger}%
\end{equation}
as%
\begin{eqnarray}
H^{(E,o)}=(J-h)\left(  2c_{0}^{\dagger}c_{0}-1\right)  -(J+h)\left(  2c_{\pi
}^{\dagger}c_{\pi}-1\right)  \nonumber\\
\quad\quad\quad\quad+\sum_{q\in q^{(E,o)},q\neq0,\pi}\omega(q)\left(
2\eta_{q}^{\dagger}\eta_{q}-1\right)  , \label{H(E,o)}%
\end{eqnarray}
where
\begin{eqnarray}
u_{q}^{2}  &  =\frac{1}{2}\left(  1+\frac{\epsilon(q)}{\omega(q)}\right)
,v_{q}^{2}=\frac{1}{2}\left(  1-\frac{\epsilon(q)}{\omega(q)}\right)
,2u_{q}v_{q}=\frac{\Delta(q)}{\omega(q)},\nonumber\\
\omega(q)  &  =\sqrt{\epsilon(q)^{2}+\Delta(q)^{2}},\epsilon(q)=J\cos
{q}-h,\Delta(q)=J\sin{q.}%
\end{eqnarray}
Notice there is no need of Bogoliubov transformation for $q=0$ and $\pi$.

\subsection{Diagonalization in the even channel}

In the even channel ($M\in even$), the Jordan-Wigner fermions must obey
anti-PBC: $c_{N+j}=-c_{j}$, and the momentum in the $1$st BZ must take a value
in the set
\begin{equation}
q^{(E,e)}=\{-\frac{N-1}{N}\pi,\ldots,-\frac{1}{N}\pi,\frac{1}{N}\pi
,\ldots,\frac{N-1}{N}\pi\}.
\end{equation}
The diagonalized Hamiltonian is%
\begin{equation}
H^{(E,e)}=\sum_{q\in q^{(E,e)}}\omega(q)\left(  2\eta_{q}^{\dagger}\eta
_{q}-1\right)  . \label{H(E,e)}%
\end{equation}

\subsection{Band structure of the energy levels}

\subsubsection{The ground state(s) and energy gap}

The lowest energy state in the odd channel is%

\begin{equation}
|E_{0}^{(E,o)}\rangle=c_{\pi}^{\dag}|\phi^{(E,o)}\rangle,
\label{state E0(E,o)}%
\end{equation}
where $|\phi^{(E,o)}\rangle$ is a pure BCS-like function,%
\begin{equation}
|\phi^{(E,o)}\rangle=\prod_{\substack{q\in q^{(E,o)},0<q<\pi}}\left(
u_{q}+\mathrm{i}v_{q}c_{q}^{\dag}c_{-q}^{\dag}\right)  |0\rangle, \label{BCS-Eo}%
\end{equation}
where the vacuum state $|0\rangle$ is (\ref{vacuum}). Its energy reads
\begin{equation}
E_{0}^{(E,o)}=\left\vert J-h\right\vert -(J-h)-\sum_{q\in q^{(E,o)}}\omega(q).
\end{equation}
Notice that the fermionic BCS state $|\phi^{(E,o)}\rangle$ itself can not be a
valid state for the original spin model because of the parity constraint.

Likewise, The lowest energy state in the even channel reads%

\begin{equation}
|E_{0}^{(E,e)}\rangle=|\phi^{(E,e)}\rangle, \label{state E0(E,e)}%
\end{equation}
where%
\begin{equation}
|\phi^{(E,e)}\rangle=\prod_{\substack{q\in q^{(E,e)},q>0}}\left(
u_{q}+\mathrm{i}v_{q}c_{q}^{\dag}c_{-q}^{\dag}\right)  |0\rangle. \label{BCS-Ee}%
\end{equation}
Its energy reads%
\begin{equation}
E_{0}^{(E,e)}=-\sum_{q\in q^{(E,e)}}\omega(q) \label{E0(E,e)}%
\end{equation}

If $N$ is small, we always have $E_{0}^{(E,e)}<E_{0}^{(E,o)}$, so
$|E_{0}^{(E,e)}\rangle$ is the ground state.

If $N\rightarrow\infty$, $|E_{0}^{(E,e)}\rangle$ is still the ground state for
$J<h$ and there is a gap, $\Delta_{gap}=2(h-J)$, to the first excited state
$|E_{0}^{(E,o)}\rangle$. Above $|E_{0}^{(E,o)}\rangle$, there is a continuum
band of excitations. While for $J>h$, $|E_{0}^{(E,e)}\rangle$ and
$|E_{0}^{(E,o)}\rangle$ become the degenerate ground states and there is a
gap, $\Delta_{gap}=4(J-h)$, above them. Now, the sum in the ground state
energy, (\ref{E0(E,e)}), can be replaced with an integral that can be worked
out, so we get
\begin{equation}
\left.  \frac{E_{0}^{(E,e)}}{N}\right\vert _{N\rightarrow\infty}%
\longrightarrow\frac{-2\left\vert J-h\right\vert }{\pi}E\left(  \frac
{-4Jh}{(J-h)^{2}}\right)  , \label{E0(E,e)/N}%
\end{equation}
where $E(x)$ is the complete elliptic integral of the second kind.
(\ref{E0(E,e)/N}) is non-analytic at $J/h=1$, because its second derivative in
respect of $J/h$ has a logarithmic divergent peak $\sim(1/\pi)\ln|J/h-1|$. So
in fact, we have a critical point at $J=h$. These conclusions are the same as
the ones in previous investigations \cite{SachdevScience}.

We have checked that the two states, (\ref{state E0(E,o)}) and
(\ref{state E0(E,e)}), in the limit $h\rightarrow0$ correspond to two GHZ spin
states in $\sigma^{x}$ representation,%
\begin{eqnarray}
 \lim_{h\rightarrow0}|E_{0}^{(E,e)}\rangle=\frac{1}{\sqrt{2}}(|\cdots
\leftarrow_{j-1},\rightarrow_{j},\leftarrow_{j+1},\rightarrow_{j+2}%
,\cdots\rangle  \nonumber\\
\quad\quad\quad\quad\quad\quad-|\cdots\rightarrow_{j-1},\leftarrow_{j},\rightarrow
_{j+1},\leftarrow_{j+2},\cdots\rangle),\\
 \lim_{h\rightarrow0}|E_{0}^{(E,o)}\rangle=\frac{1}{\sqrt{2}}(|\cdots
\leftarrow_{j-1},\rightarrow_{j},\leftarrow_{j+1},\rightarrow_{j+2}%
,\cdots\rangle \nonumber\\
\quad\quad\quad\quad\quad\quad+|\cdots\rightarrow_{j-1},\leftarrow_{j},\rightarrow
_{j+1},\leftarrow_{j+2},\cdots\rangle),
\end{eqnarray}
respectively.

\subsubsection{Analysis of bands}

\label{332}

The degrees of freedom (DOF) of the fermionic problem are $2^{N}$ for both
channels, so we get $2^{N+1}$ DOF totally, which is redundantly twice of the
DOF of the original spin model. However, the odd channel requires an odd
parity and the even channel an even parity. This parity constraint helps us to
obliterate the redundant DOF in each channel exactly and reconstruct the band
structure of the original spin problem.

We can construct all excited energy levels by the BCS functions (\ref{BCS-Ee})
and (\ref{BCS-Eo}) precisely. All energy levels can be grouped into bands that
are labelled by a set of indexes $(P,Q^{(E)},n_{0},n_{\pi})$, where $P$ is the
parity defined in (\ref{P}), $Q^{(E)}$ is a quasi-particle number defined as
\begin{equation}
Q^{(E)}=\sum_{q\in q^{(E,o)}\cup q^{(E,e)}}n_{q}, \label{Q}%
\end{equation}
$n_{0}=c_{0}^{\dagger}c_{0}$, and $n_{\pi}=c_{\pi}^{\dagger}c_{\pi}$. For
example, the band indexes of the two states discussed above are $(1,0,0,0)$
and $(-1,1,0,1)$ for $|E_{0}^{(E,e)}\rangle$ and $|E_{0}^{(E,o)}\rangle$
respectively. These two bands contains only one level each. From all bands of
the fermionic problem, we can pick out the valid ones for the spin model
according to the parity constraint. Several valid bands of low energy states
are listed in Table \ref{table:E}. The energy value of each state is readily
read out from the diagonalized Hamiltonian, (\ref{H(E,o)}) and (\ref{H(E,e)}).
The band structure is available for arbitrary $N$ ($N\in Even$, $2\leq
N<\infty$). It is noteworthy that the invalid bands, for example such as
$(1,1,0,0)$ and $(-1,2,0,1)$, are prohibited by the parity constraint of each channel.

\begin{table}[ptb]
\begin{center}%
\begin{tabular}
[c]{|c|c|c|}\hline
Valid Bands & Fermionic states & Number\\
$(P,Q^{(E)},\eta_{0},\eta_{\pi})$ & $(q\neq0,\pi)$ & of states\\\hline
\multicolumn{3}{|c|}{odd channel ($M\in odd$)}\\\hline
$(-1,1,0,1)$ & $c_{\pi}^{\dagger}|\phi^{\left(  E,o\right)  }\rangle
=|E_{0}^{\left(  E,o\right)  }\rangle$ & $1$\\\hline
$(-1,1,1,0)$ & $c_{0}^{\dagger}|\phi^{\left(  E,o\right)  }\rangle$ &
$1$\\\hline
$(-1,1,0,0)$ & $\eta_{q}^{\dagger}|\phi^{\left(  E,o\right)  }\rangle$ &
$\mathrm{C}_{N-2}^{1}$\\\hline
$(-1,3,0,0)$ & $\eta_{q_{1}}^{\dagger}\eta_{q_{2}}^{\dagger}\eta_{q_{3}%
}^{\dagger}|\phi^{\left(  E,o\right)  }\rangle$ & $\mathrm{C}_{N-2}^{3}%
$\\\hline
$(-1,3,0,1)$ & $\eta_{q_{1}}^{\dagger}\eta_{q_{2}}^{\dagger}c_{\pi}^{\dagger
}|\phi^{\left(  E,o\right)  }\rangle$ & $\mathrm{C}_{N-2}^{2}$\\\hline
$(-1,3,1,0)$ & $\eta_{q_{1}}^{\dagger}\eta_{q_{2}}^{\dagger}c_{0}^{\dagger
}|\phi^{\left(  E,o\right)  }\rangle$ & $\mathrm{C}_{N-2}^{2}$\\\hline
$(-1,3,1,1)$ & $\eta_{q_{1}}^{\dagger}c_{0}^{\dagger}c_{\pi}^{\dagger}%
|\phi^{\left(  E,o\right)  }\rangle$ & $\mathrm{C}_{N-2}^{1}$\\\hline
$\vdots$ & $\vdots$ & $\vdots$\\\hline
\multicolumn{3}{|c|}{even channel ($M\in even$)}\\\hline
$(1,0,0,0)$ & $|\phi^{\left(  E,e\right)  }\rangle=|E_{0}^{\left(  E,e\right)
}\rangle$ & $1$\\\hline
$(1,2,0,0)$ & $\eta_{q_{1}}^{\dagger}\eta_{q_{2}}^{\dagger}|\phi^{\left(
E,e\right)  }\rangle$ & $\mathrm{C}_{N}^{2}$\\\hline
$(1,4,0,0)$ & $\eta_{q_{1}}^{\dagger}\eta_{q_{2}}^{\dagger}\eta_{q_{3}%
}^{\dagger}\eta_{q_{4}}^{\dagger}|\phi^{\left(  E,e\right)  }\rangle$ &
$\mathrm{C}_{N}^{4}$\\\hline
$\vdots$ & $\vdots$ & $\vdots$\\\hline
\end{tabular}
\end{center}
\caption{The valid bands satisfying the odd or even parity constraint for the
case $N\in Even$. Invalid bands are not included. The energy value of each
state is readily read out from the diagonalized Hamiltonian, (\ref{H(E,o)}) or
(\ref{H(E,e)}).}%
\label{table:E}%
\end{table}

%%%%%%%%%%% FIG 2  %%%%%%%%%%%%%%%%%%%%%%%%%%%%%%%
\begin{figure}[ptb]
\begin{center}
\includegraphics[width=3.2in,angle=0]{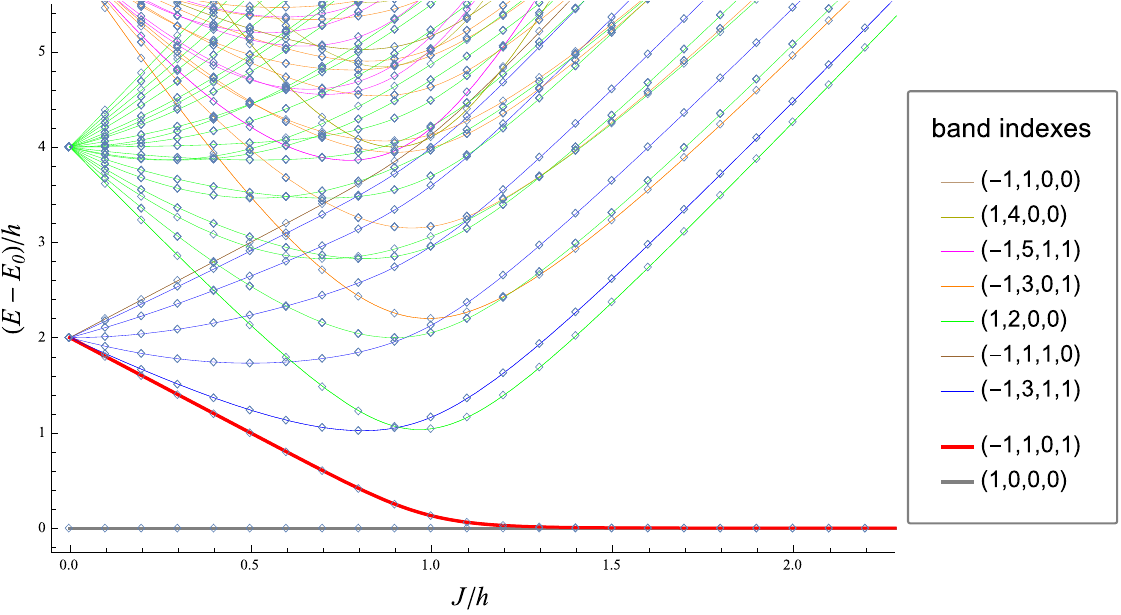}
\end{center}
\caption{The band structure of low-energy levels for a system of $N=12$. The
purpose of this figure is to check the band structure for $N\in Even$
disclosed in the text. The dingbat diamonds ("$\diamond$") denote the results
from exact diagonalization on the original spin model, which are in perfect
coincidence with the bands of levels. The true ground state,
(\ref{state E0(E,e)}), with band indexes $(1,0,0,0)$ is set as a reference.}%
\label{N=12}%
\end{figure}
%%%%%%%%%%% FIG 2  %%%%%%%%%%%%%%%%%%%%%%%%%%%%%%%

%%%%%%%%%%% FIG 3  %%%%%%%%%%%%%%%%%%%%%%%%%%%%%%%
\begin{figure}[ptb]
\begin{center}
\includegraphics[width=3.2in,angle=0]{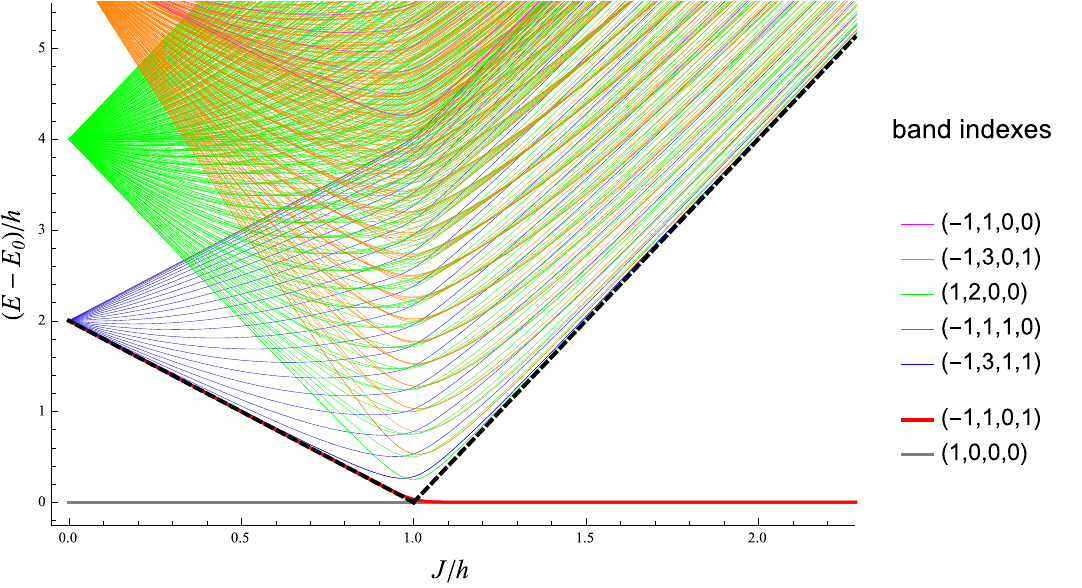}
\end{center}
\caption{The band structure of low-energy levels for a system of $N=50$. From
this figure, one can figure out the trend for $N\rightarrow\infty$\ (we still
hold $N\in Even$). The dashed black line is the lower bound of excitations
when $N\rightarrow\infty$, whose intersecting point at $J/h=1$ is a critical
point as disclosed by (\ref{E0(E,e)}). The bottom levels of many bands will
touch this critical point, which results in a divergent DOS. At both sides of
the critical point, the system is gapped. Not all bands above the dashed black
line are shown.}%
\label{N=50}%
\end{figure}
%%%%%%%%%%% FIG 3  %%%%%%%%%%%%%%%%%%%%%%%%%%%%%%%

%\paragraph{Cases $N=12$, $N=50$, and $N(\in Even)\rightarrow\infty$}

To testify the band structure further, we compare it with the result by an
exact diagonalization of the transverse Ising model with a small size, say
$N=12$. The comparison is shown in Fig. 2, where a perfect coincidence can be
clearly seen. So we see our method can restore the full degrees of freedom of
the spin model.

The band structure for a larger system, say $N=50$, is exemplified in Fig. 3,
through which we can see the trend for a large enough system, $N\rightarrow
\infty$. At the critical point, the bottom levels of many bands satisfying
$P=1$ or $P\times n_{\pi}=-1$ will touch the critical point, which will result
in a divergent DOS.

\section{A-cycle problem with ring frustration ($N\in Odd$)}

\label{sec:odd}

Now we turn to the interesting case with frustration (Fig. \ref{fig1}). The
procedure is almost the same. But the story is totally different. In the
strong antiferromagnetic region, we find a gapless spectrum above the ground
state if the system is large enough.

\subsection{Diagonalization in the odd channel}

In the odd channel ($M\in odd$), the Jordan-Wigner fermions must obey PBC:
$c_{N+j}=c_{j}$, and the momentum in the $1$st BZ must take a value in the
set
\begin{equation}
q^{(O,o)}=\{-\frac{N-1}{N}\pi,\ldots,-\frac{2}{N}\pi,0,\frac{2}{N}\pi
,\ldots,\frac{N-1}{N}\pi\}.
\end{equation}
The diagonalized Hamiltonian is%
\begin{equation}
H^{(O,o)}=(J-h)\left(  2c_{0}^{\dagger}c_{0}-1\right)  +\sum_{q\in
q^{(O,o)},q\neq0}\omega(q)\left(  2\eta_{q}^{\dagger}\eta_{q}-1\right)  .
\label{H(O,o)}%
\end{equation}

\subsection{Diagonalization in the even channel}

In the even channel ($M\in even$), the Jordan-Wigner fermions must obey
anti-PBC: $c_{N+j}=-c_{j}$, and the momentum in the $1$st BZ must take a value
in the set
\begin{equation}
q^{(O,e)}=\{-\frac{N-2}{N}\pi,\ldots,-\frac{1}{N}\pi,\frac{1}{N}\pi
,\ldots,\frac{N-2}{N}\pi,\pi\}.
\end{equation}
The diagonalized Hamiltonian is%
\begin{equation}
H^{(O,e)}=-(J+h)\left(  2c_{\pi}^{\dagger}c_{\pi}-1\right)  +\sum_{q\in
q^{(O,e)},q\neq\pi}\omega(q)\left(  2\eta_{q}^{\dagger}\eta_{q}-1\right)  .
\label{H(O,e)}%
\end{equation}

\subsection{Band structure of the energy levels}

\subsubsection{The ground state}

The lowest energy state in the odd channel reads%

\begin{equation}
|E_{0}^{(O,o)}\rangle=c_{0}^{\dag}|\phi^{(O,o)}\rangle, \label{state E0(O,o)}%
\end{equation}
where%
\begin{equation}
|\phi^{(O,o)}\rangle=\prod_{\substack{q\in q^{(O,o)},q>0}}\left(
u_{q}+\mathrm{i}v_{q}c_{q}^{\dag}c_{-q}^{\dag}\right)  |0\rangle. \label{BCS-Oo}%
\end{equation}
Its energy reads
\begin{equation}
E_{0}^{(O,o)}=\left\vert J-h\right\vert +(J-h)-\sum_{q\in q^{(O,o)}}\omega(q).
\end{equation}

The lowest energy state in the even channel reads%

\begin{equation}
|E_{0}^{(O,e)}\rangle=\eta_{\frac{\pi}{N}}^{\dag}c_{\pi}^{\dag}|\phi
^{(O,e)}\rangle, \label{state E0(O,e)}%
\end{equation}
where%
\begin{equation}
|\phi^{(O,e)}\rangle=\prod_{\substack{q\in q^{(O,e)},0<q<\pi}}\left(
u_{q}+\mathrm{i}v_{q}c_{q}^{\dag}c_{-q}^{\dag}\right)  |0\rangle. \label{BCS-Oe}%
\end{equation}
Its energy reads%
\begin{equation}
E_{0}^{(O,e)}=2\omega(\frac{\pi}{N})-\sum_{q\in q^{(O,e)}}\omega(q).
\end{equation}

If $N$ is small, we always have $E_{0}^{(O,o)}<E_{0}^{(O,e)}$, so
$|E_{0}^{(O,o)}\rangle$ is the ground state and $|E_{0}^{(O,e)}\rangle$ is the
first excited state.

If $N\rightarrow\infty$, the state $|E_{0}^{(O,o)}\rangle$
(\ref{state E0(O,o)}) is the ground state. In the region $J<h$, there is a
gap, $\Delta_{gap}=2(h-J)$, to the first excited state $|E_{0}^{(O,e)}\rangle$
(\ref{state E0(O,e)}). In the region $J>h$, the energy gap between
$|E_{0}^{(O,o)}\rangle$ and $|E_{0}^{(O,e)}\rangle$ disappears. But we notice
that there is no energy gap between $|E_{0}^{(O,e)}\rangle$ and the next
excitation, and so forth. In fact there appears a gapless spectrum above the
ground state $|E_{0}^{(O,o)}\rangle$. We will discuss this gapless spectrum
later in detail in Section \ref{gapless spectrum}. While at $J=h$, the ground
state energy,%
\begin{equation}
\left.  \frac{E_{0}^{(O,o)}}{N}\right\vert _{N\rightarrow\infty}%
\longrightarrow\frac{-2\left\vert J-h\right\vert }{\pi}E\left(  \frac
{-4Jh}{(J-h)^{2}}\right)  +\frac{2}{N}(J-h)\theta(J-h), \label{E0(O,o)/N}%
\end{equation}
with a Heaviside step function $\theta(x)$ is still non-analytic. In fact, the
self-duality still holds for the frustrated ring system with odd $N$ and
ensures the ocurring of quantum phase transition at $J=h$. One can see this
clear by defining new Ising-type operators,
\begin{equation}
\tau_{j}^{z}=-\sigma_{j}^{x}\sigma_{j+1}^{x},\tau_{j}^{x}=(-1)^{j}\prod
_{l<j}\sigma_{l}^{z},
\end{equation}
to get a dual form of Hamiltonian \cite{Kogut}%
\begin{equation}
H=-J\sum_{j=1}^{N}\tau_{j}^{z}+h\sum_{j=1}^{N}\tau_{j}^{x}\tau_{j+1}^{x}.
\end{equation}

\subsubsection{Analysis of bands}

By defining the quasi-particle number
\begin{equation}
Q^{(O)}=\sum_{q\in q^{(O,o)}\cup q^{(O,e)}}n_{q},
\end{equation}
we can use the set of indexes $(P,Q^{(O)},n_{0},n_{\pi})$ to label all the
fermionic bands as we have done in Section \ref{332}. From the fermionic bands
in each channel, we can pick out the valid ones for the original spin model
according to the parity constraint. The valid bands of several low energy
levels are listed in Table \ref{table:O}. The energy value of each state is
readily read out from the diagonalized Hamiltonian, (\ref{H(O,o)}) or
(\ref{H(O,e)}). The band structure is available for arbitrary $N$ ($N\in Odd$,
$3\leq N<\infty$).

%%%%%%%%%%% Table 2  %%%%%%%%%%%%%%%%%%%%%%%%%%%%%%%
\begin{table}[ptb]
\begin{center}%
\begin{tabular}
[c]{|c|c|c|}\hline
Valid Bands & Fermionic state & Number\\
$(P,Q^{(O)},\eta_{0},\eta_{\pi})$ & $(q\neq0,\pi)$ & of states\\\hline
\multicolumn{3}{|c|}{odd channel ($M\in odd$)}\\\hline
$(-1,1,0,0)$ & $\eta_{q}^{\dagger}|\phi^{\left(  O,o\right)  }\rangle$ &
$\mathrm{C}_{N-1}^{1}$\\\hline
$(-1,1,1,0)$ & $c_{0}^{\dagger}|\phi^{\left(  O,o\right)  }\rangle
=|E_{0}^{\left(  O,o\right)  }\rangle$ & $1$\\\hline
$(-1,3,0,0)$ & $\eta_{q_{1}}^{\dagger}\eta_{q_{2}}^{\dagger}\eta_{q_{3}%
}^{\dagger}|\phi^{\left(  O,o\right)  }\rangle$ & $\mathrm{C}_{N-1}^{3}%
$\\\hline
$(-1,3,1,0)$ & $\eta_{q_{1}}^{\dagger}\eta_{q_{2}}^{\dagger}c_{0}^{\dagger
}|\phi^{\left(  O,o\right)  }\rangle$ & $\mathrm{C}_{N-1}^{2}$\\\hline
$\vdots$ & $\vdots$ & $\vdots$\\\hline
\multicolumn{3}{|c|}{even channel ($M\in even$)}\\\hline
$(1,0,0,0)$ & $|\phi^{\left(  O,e\right)  }\rangle$ & $1$\\\hline
$(1,2,0,0)$ & $\eta_{q_{1}}^{\dagger}\eta_{q_{2}}^{\dagger}|\phi^{\left(
O,e\right)  }\rangle$ & $\mathrm{C}_{N-1}^{2}$\\\hline
$(1,2,0,1)$ & $\eta_{q}^{\dagger}c_{\pi}^{\dagger}|\phi^{\left(  O,e\right)
}\rangle$ & $\mathrm{C}_{N-1}^{1}$\\\hline
$\vdots$ & $\vdots$ & $\vdots$\\\hline
\end{tabular}
\end{center}
\caption{The valid bands satisfying the odd or even parity constraint for the
case $N\in Odd$. The energy value of each state is readily read out from the
diagonalized Hamiltonian, (\ref{H(O,o)}) or (\ref{H(O,e)}).}%
\label{table:O}%
\end{table}
%%%%%%%%%%% Table 2  %%%%%%%%%%%%%%%%%%%%%%%%%%%%%%%

%%%%%%%%%%% FIG 4  %%%%%%%%%%%%%%%%%%%%%%%%%%%%%%%
\begin{figure}[ptb]
\begin{center}
\includegraphics[width=0.7\textwidth]{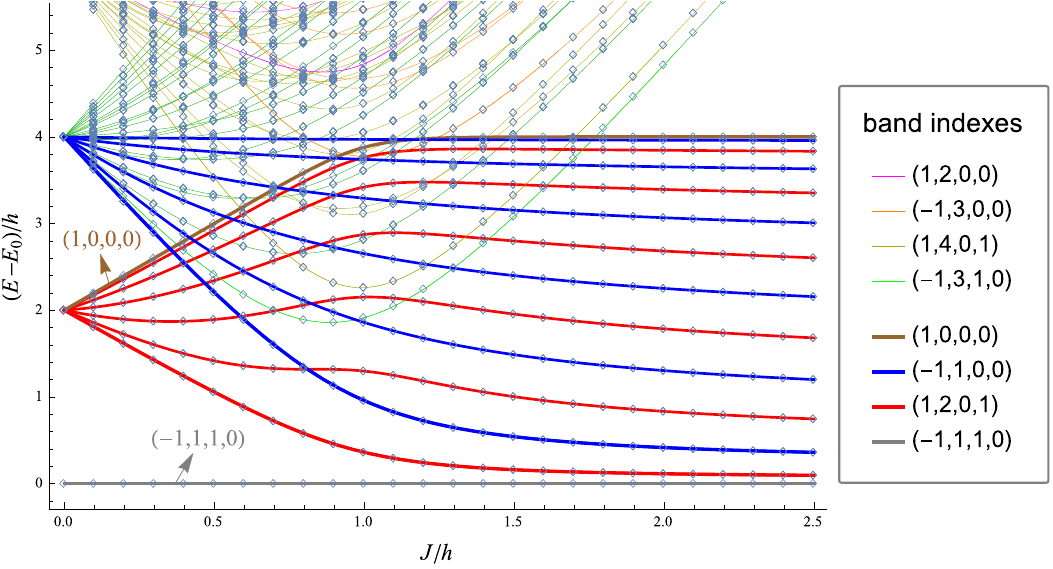}
\end{center}
\caption{The band structure of low-energy levels for a system of $N=13$. The
purpose of this figure is to check the band structure for $N\in Odd$ disclosed
in the text. The bands of levels are in perfect coincidence with the results
from exact diagonalization on the original spin model (shown as dingbat
diamonds "$\diamond$"). The true ground state, (\ref{state E0(O,o)}), with
band indexes $(-1,1,1,0)$ is set as a reference.}%
\label{N=13}%
\end{figure}
%%%%%%%%%%% FIG 4  %%%%%%%%%%%%%%%%%%%%%%%%%%%%%%%

%%%%%%%%%%% FIG 5  %%%%%%%%%%%%%%%%%%%%%%%%%%%%%%%
\begin{figure}[ptb]
\begin{center}
\includegraphics[width=0.7\textwidth]{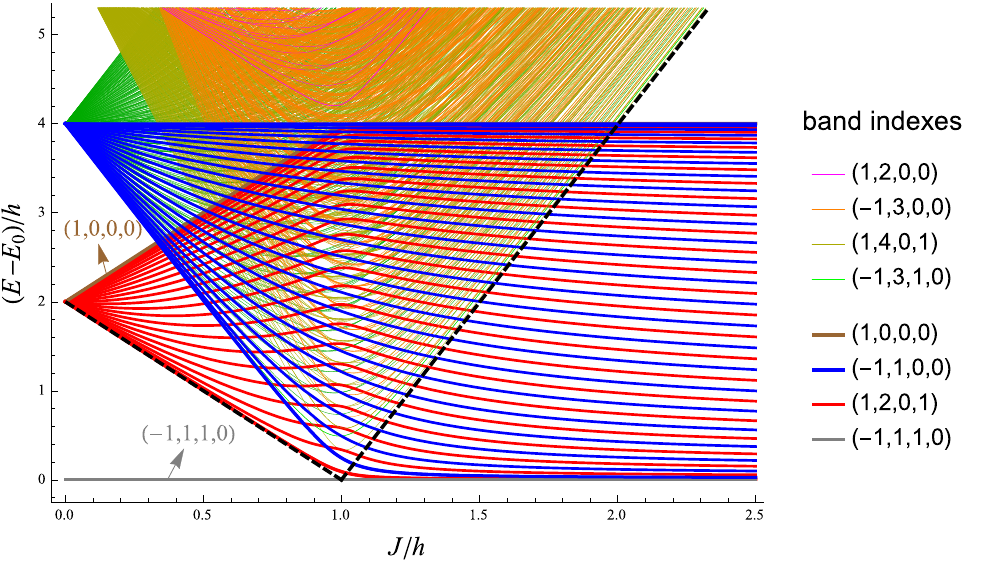}
\end{center}
\caption{The band structure of low-energy levels for a system of $N=51$. From
this figure, one can figure out the trend for $N\rightarrow\infty$\ ($N\in
Odd$ is still hold). In contrast with the case of $N\in Even$ in Fig.
\ref{N=50}, the low excitations is dramatically changed in the region $J/h>1$
due to the presence of ring frustration. If $N\rightarrow\infty$, we can still
draw a dashed black line, whose intersecting point at $J/h=1$ is a critical
point as disclosed by (\ref{E0(O,o)/N}). And the bottom levels of many bands
satisfying $P=-1$ or $P\times n_{\pi}=1$ will touch this critical point, which
results in a divergent DOS. There is an energy gap for the region $J/h<1$.
However, the excitations for the region $J/h>1$ are gapless, where the energy
interval of width $4h$ is depleted by $N+1$ levels occupied by $2N$ states
involving 4 bands: $(-1,1,1,0)$, $(1,2,0,1)$, $(-1,1,0,0)$, and $(1,0,0,0)$.
Not all bands above the dashed black line are shown.}%
\label{N=51}%
\end{figure}
%%%%%%%%%%% FIG 5  %%%%%%%%%%%%%%%%%%%%%%%%%%%%%%%

%\paragraph{Cases $N=13$, $N=51$, and $N(\in Odd)\rightarrow\infty$}

In Fig. 4, we testify the band structure further by comparing it with the
result by the exact diagonalization on a system of $N=13$. Perfect coincidence
is observed. So we see our method restores the full degrees of freedom of the
spin model.

In Fig. 5, the band structure for a larger system with $N=51$ is exemplified,
through which we can see the trend for a large enough system, $N\rightarrow
\infty$. At the critical point, the bottom levels of many bands satisfying
$P=-1$ or $P\times n_{\pi}=1$ will touch the critical point, which will result
in a divergent DOS.

\subsubsection{ Gapless spectrum in the region $J>h$}

\label{gapless spectrum}

In the strong antiferromagnetic region $J>h$, there forms a gapless spectrum
when $N\rightarrow\infty$. It contains $2N$ states involving $4$ interwoven
bands. They occupy $N+1$ energy levels. The ground state $|E_{0}%
^{(O,o)}\rangle$ with band indexes $(-1,1,1,0)$ lies at the bottom. We relabel
it as
\begin{equation}
|E_{0}\rangle=|E_{0}^{(O,o)}\rangle=c_{0}^{\dag}|\phi^{(O,o)}\rangle.
\label{state E0}%
\end{equation}
The upper-most state is $|\phi^{(O,e)}\rangle$ with indexes $(1,0,0,0)$. We
relabel it as
\begin{equation}
|E_{\pi}\rangle=|\phi^{(O,e)}\rangle. \label{state Epi}%
\end{equation}
The other two bands are:
\begin{equation}
|E_{q}\rangle=\eta_{q}^{\dag}c_{\pi}^{\dag}|\phi^{(O,e)}\rangle,(q\in
q^{(O,e)},q\neq\pi) \label{state EOe}%
\end{equation}
with indexes $(1,2,0,1)$ and
\begin{equation}
|E_{q}\rangle=\eta_{q}^{\dag}|\phi^{(O,o)}\rangle,(q\in q^{(O,o)},q\neq0)
\label{state EOo}%
\end{equation}
with indexes $(-1,1,0,0)$. If $N$ is finite, $|E_{0}\rangle$ and $|E_{\pi
}\rangle$ are nondegenerate, while other $|E_{q}\rangle$'s are doubly
degenerate. When $N\rightarrow\infty$, these $2N$ states deplete the energy
interval of width
\begin{equation}
\Delta_{w}=E_{\pi}-E_{0}\overset{N\rightarrow\infty}{\longrightarrow}4h
\end{equation}
between $|E_{0}\rangle$ and $|E_{\pi}\rangle$. This result is beyond the
familiar schematic picture for quantum phase transition \cite{SachdevScience}.

\paragraph{Perturbative theory}

\begin{figure}[ptb]
\begin{center}
\includegraphics[width=0.7\textwidth]{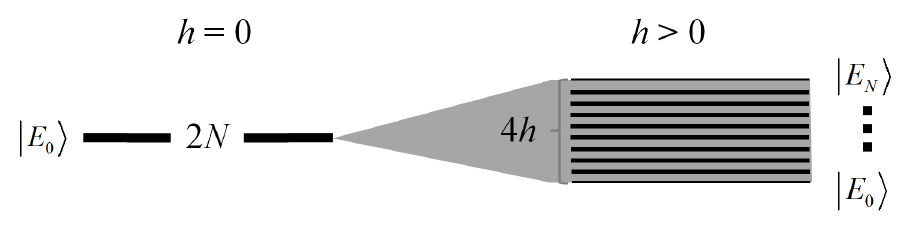}
\end{center}
\caption{Schematic diagram of the formation of gapless specrum above the
ground state from a perturbative point of view.}%
\label{perturbative pic}%
\end{figure}

To understand the formation of gapless spectrum, let us see a perturbative
picture in the strong antiferromagnetic region $J>h$. The first term of
(\ref{H1})%
\begin{equation}
H_{0}=J\sum_{j=1}^{N}\sigma_{j}^{x}\sigma_{j+1}^{x}%
\end{equation}
is a classical Ising model, whose ground states are highly degenerate as an
effect of antiferromagnetic seam \cite{McCoy Wu}. By choosing the
representation of $\sigma^{z}$, i.e. $\sigma_{j}^{z}\left\vert \uparrow
_{j}\right\rangle $=$\left\vert \uparrow_{j}\right\rangle $ and $\sigma
_{j}^{z}\left\vert \downarrow_{j}\right\rangle $=$-\left\vert \downarrow
_{j}\right\rangle $, and denoting the two eigenstates of $\sigma_{j}^{x}$ as
$\left\vert \rightarrow_{j}\right\rangle $=$(\left\vert \uparrow
_{j}\right\rangle $+$\left\vert \downarrow_{j}\right\rangle )/\sqrt{2}$ and
$\left\vert \leftarrow_{j}\right\rangle $=$(\left\vert \uparrow_{j}%
\right\rangle -\left\vert \downarrow_{j}\right\rangle )/\sqrt{2}$
\cite{Sachdev}, we can express its $2N$-fold degenerate ground states as kink
states \cite{Solomon}:
\begin{equation}
\eqalign{|K(j),\rightarrow\rangle=|\cdots,\leftarrow
_{j-1},\rightarrow_{j},\rightarrow_{j+1},\leftarrow_{j+2},\cdots
\rangle \cr
|K(j),\leftarrow\rangle=|\cdots,\rightarrow_{j-1},\leftarrow
_{j},\leftarrow_{j+1},\rightarrow_{j+2},\cdots\rangle.}
\label{kinkstates}
\end{equation}
where kinks occur between sites $j$ and $j$+$1$. The classical Ising system
falls into one of these states by spontaneous symmetry breaking \cite{Diep}.
But they are not eigenstates of the full quantum system. The second term of
(\ref{H1}),%
\begin{equation}
V=-h\sum_{j=1}^{N}\sigma_{j}^{z},
\end{equation}
as a source of quantum fluctuation, plays the role of perturbation when $h/J$
is small. We relabel the kink states as
\begin{equation}%
\eqalign{
|2j-1\rangle=|K(j),\rightarrow\rangle \cr
|2j\rangle=|K(j),\leftarrow\rangle.
}
\end{equation}
Then by the simplest perturbative scheme based on these levels, we can deduce
the matrix form of the full spin Hamiltonian $H=H_{0}+V$ ($2N\times2N$), whose
diagonal elements read
\begin{equation}
H_{2j-1,2j-1}=H_{2j,2j}=J
\end{equation}
and off-diagonal elements read%
\begin{equation}
H_{2j,2j+1}=H_{2j-1,2j+2}=H_{2j+1,2j}=H_{2j+2,2j-1}=-h.\nonumber
\end{equation}
Other elements are zero. We can arrive at an effective Hamiltonian
approximately,%
\begin{eqnarray}
H &  \approx H_{\mathrm{eff}}=J\sum_{j=1}^{N}\left(  |2j-1\rangle\left\langle
2j-1\right\vert +|2j\rangle\left\langle 2j\right\vert \right)  \nonumber\\
&  -h\sum_{j=1}^{N}\left(  |2j\rangle\left\langle 2j+1\right\vert
+|2j-1\rangle\left\langle 2j+2\right\vert +\mathrm{h.c.}\right)  .
\end{eqnarray}
Now by introducing a Fourier transformation%
\begin{equation}
|2j-1\rangle=\frac{1}{\sqrt{N}}\sum_{k}|a_{k}\rangle\mathrm{e}%
^{\mathrm{i}k\,j},|2j\rangle=\frac{1}{\sqrt{N}}\sum_{k}|b_{k}%
\rangle\mathrm{e}^{\mathrm{i}k\,j}%
\end{equation}
with%
\begin{equation}
k=-\frac{N-1}{N}\pi,\cdots,-\frac{2}{N}\pi,0,\frac{2}{N}\pi,\cdots,\frac
{N-1}{N}\pi,
\end{equation}
we get%
\begin{equation}
\fl H_{\mathrm{eff}}=\sum_{k}[J\left(  |a_{k}\rangle\left\langle a_{k}\right\vert
+|b_{k}\rangle\left\langle b_{k}\right\vert \right)  -2h\left(  \cos
k|b_{k}\rangle\left\langle a_{k}\right\vert +\cos k|a_{k}\rangle\left\langle
b_{k}\right\vert \right)  ].
\end{equation}
We can diagonalize it as%
\begin{equation}
H_{\mathrm{eff}}=\sum_{k}[\left(  J-2h\cos k\right)  |A_{k}\rangle\left\langle
A_{k}\right\vert +\left(  J+2h\cos k\right)  |B_{k}\rangle\left\langle
B_{k}\right\vert ]
\end{equation}
by denoting%
\begin{equation}
|A_{k}\rangle=\frac{1}{\sqrt{2}}\left(  |a_{k}\rangle+|b_{k}\rangle\right)
,|B_{k}\rangle=\frac{1}{\sqrt{2}}\left(  -|a_{k}\rangle+|b_{k}\rangle\right)
.
\end{equation}
Thus the degenerate ground states of $H_{0}$ is dispersed by $V$ and form a
band of $N+1$ levels, i.e. the degeneracy is partly lifted. It is easy to
check that the states $|A_{k}\rangle$ have odd parity and the states
$|B_{k}\rangle$ have even parity. If $N\rightarrow\infty$, they form a gapless
spectrum of width $4h$. They are good approximations of the lowest $2N$
rigorous energy states. For example, the ground state of the system is%
\begin{equation}
|A_{0}\rangle=\frac{1}{\sqrt{2N}}{\sum\nolimits_{j,\tau}}|K(j),\tau
\rangle\label{sum of kink states}%
\end{equation}
approximately. It is highly entangled. The excited eigenstates are
recombinations of the $2N$ kink states likewise. If $N\rightarrow\infty$, the
low-lying excitations form a gapless spectrum of width $4h$ as the schematic
plot in Fig. \ref{perturbative pic}.

%%%%%%%%%%% FIG 6  %%%%%%%%%%%%%%%%%%%%%%%%%%%%%%%
%%%%%%%%%%% FIG 6  %%%%%%%%%%%%%%%%%%%%%%%%%%%%%%%
As a relevant issue, we found that, if one deduces an effective
two-dimensional classical Ising model for the quantum Ising chain by the
first-order Suzuki-Trotter decomposition in the usual way
\cite{Suzuki,MSuzuki}, the model will fail to capture the lifting of
degeneracy of the kink states.

\subsection{Correlation function of the ground state}

Now we concern the longitudinal correlation function of the ground state. We
still follow the strategy: try to work out the correlation function as a
function of $N\in Odd$, then set the limit, $N\rightarrow\infty$, to see if
there is any surprising result. For the gapless region, we find a new type of
Toeplitz determinant that needs to be evaluated rigorously.

The two-point longitudinal spin-spin correlation function of the ground state
is defined as
\begin{equation}
C_{r,N}^{xx}=\langle\phi^{(O,o)}|c_{0}\sigma_{j}^{x}\sigma_{j+r}^{x}%
c_{0}^{\dag}|\phi^{(O,o)}\rangle.
\end{equation}
By introducing the operators, $A_{j}$=$c_{j}^{\dag}$+$c_{j}$ and $B_{j}%
$=$c_{j}^{\dag}-c_{j}$, with the relations, $A_{j}^{2}$=$1$ and $A_{j}B_{j}%
$=$\exp(-$i$\pi c_{j}^{\dag}c_{j})$, we get
\begin{equation}
C_{r,N}^{xx}=\langle\phi^{(O,o)}|c_{0}B_{j}A_{j+1}B_{j+1}\ldots B_{j+r-1}%
A_{j+r}c_{0}^{\dag}|\phi^{(O,o)}\rangle. \label{Cxx}%
\end{equation}
By making use of the Wick's theorem and the contractions in respect of
$|\phi^{(O,o)}\rangle$: $\langle c_{0}c_{0}^{\dag}\rangle$=$1$, $\langle
A_{j}c_{0}^{\dag}\rangle$=$-\langle B_{j}c_{0}^{\dag}\rangle$=$\frac{1}%
{\sqrt{N}}$, $\langle A_{j}A_{j+r}\rangle$=$-\langle B_{j}B_{j+r}\rangle
$=$\delta_{r,0}$, and $\langle B_{j}A_{j+r}\rangle$= $\mathscr{D}_{r+1}$ with
\begin{equation}
\mathscr{D}_{r}=\frac{1}{N}\sum_{_{\substack{q\in q^{\left(  O,o\right)
},q\neq0}}}\exp\left(  -\mathrm{i}qr\right)  D(\mathrm{e}^{\mathrm{i}%
q})-\frac{1}{N} \label{Dr}%
\end{equation}
where
\begin{equation}
D(\mathrm{e}^{\mathrm{i}q})=-\frac{J-h\mathrm{e}^{\mathrm{i}q}}%
{\sqrt{\left(  J-h\mathrm{e}^{\mathrm{i}q}\right)  \left(
J-h\mathrm{e}^{-\mathrm{i}q}\right)  }}.
\end{equation}
We arrive at a Toeplitz determinant%
\begin{equation}
C_{r,N}^{xx}=\left\vert
\begin{array}
[c]{cccc}%
\mathscr{D}_{0}+\frac{2}{N} & \mathscr{D}_{-1}+\frac{2}{N} & \cdots &
\mathscr{D}_{-r+1}+\frac{2}{N}\\
\mathscr{D}_{1}+\frac{2}{N} & \mathscr{D}_{0}+\frac{2}{N} & \cdots &
\mathscr{D}_{-r+2}+\frac{2}{N}\\
\cdots & \cdots & \cdots & \cdots\\
\mathscr{D}_{r-1}+\frac{2}{N} & \mathscr{D}_{r-2}+\frac{2}{N} & \cdots &
\mathscr{D}_{0}+\frac{2}{N}%
\end{array}
\right\vert . \label{Cxxdet}%
\end{equation}
It can be evaluated for arbitrary $r$ and $N$ directly. Notice that
$C_{r,N}^{xx}=C_{N-r,N}^{xx}$ due to the ring geometry. Next, we define%
\begin{equation}
D_{r}=\frac{1}{N}\sum_{_{\substack{q\in q^{\left(  O,o\right)  }}}}\exp\left(
-\mathrm{i}qr\right)  D(\mathrm{e}^{\mathrm{i}q}) \label{Drgapless}%
\end{equation}
with appropriate predefined $D(\mathrm{e}^{\mathrm{i}0})$. Thus we have%
\begin{equation}
\mathscr{D}_{r}=D_{r}-\frac{D(\mathrm{e}^{\mathrm{i}0})}{N}-\frac{1}{N}%
\end{equation}

In the gapped region ($J$ $<h$), we have $D(\mathrm{e}^{\mathrm{i}0})=1$
and $\mathscr{D}_{r}=D_{r}-\frac{2}{N}$. The correlation function is given by%
\begin{equation}
C_{r,N}^{xx}=\left\vert
\begin{array}
[c]{cccc}%
D_{0} & D_{-1} & \cdots & D_{-r+1}\\
D_{1} & D_{0} & \cdots & D_{-r+2}\\
\cdots & \cdots & \cdots & \cdots\\
D_{r-1} & D_{r-2} & \cdots & D_{0}%
\end{array}
\right\vert .
\end{equation}
This is the conventional Toeplitz determinant that has been investigated in
the previous works \cite{McCoy Wu}, the correlation function decays
exponentially with a finite correlation length $\xi=-1/\ln\left(  J/h\right)
$.

While in our focused gapless region ($J/h$ $>$ $1$), we have
$D(\mathrm{e}^{\mathrm{i}0})=-1$ and $\mathscr{D}_{r}=D_{r}$. Then the
correlation function is given by%
\begin{equation}
C_{r,N}^{xx}=\Theta(r,N)=\left\vert
\begin{array}
[c]{cccc}%
D_{0}+\frac{2}{N} & D_{-1}+\frac{2}{N} & \cdots & D_{-r+1}+\frac{2}{N}\\
D_{1}+\frac{2}{N} & D_{0}+\frac{2}{N} & \cdots & D_{-r+2}+\frac{2}{N}\\
\cdots & \cdots & \cdots & \cdots\\
D_{r-1}+\frac{2}{N} & D_{r-2}+\frac{2}{N} & \cdots & D_{0}+\frac{2}{N}%
\end{array}
\right\vert \label{Cxxdet1}%
\end{equation}
The extra term $\frac{2}{N}$ in each element makes it a totally new Toepolitz
determinant. If one erases the term $\frac{2}{N}$ when taking the limit
$N\rightarrow\infty$, the conventional Toeplitz determinant is arrived. But we
will show its non-local information is omitted in doing so. To this purpose,
we retain the term $\frac{2}{N}$ and keep $N$ as a variable. Another reason
for retaining the term $\frac{2}{N}$ is the fact that the dimension of the
determinant is $r\times r$, which can lead to a total contribution
proportional to $\frac{r}{N}$.

%%%%%%%%%%% FIG 7  %%%%%%%%%%%%%%%%%%%%%%%%%%%%%%%
\begin{figure}[ptb]
\begin{center}
\includegraphics[width=0.7\textwidth]{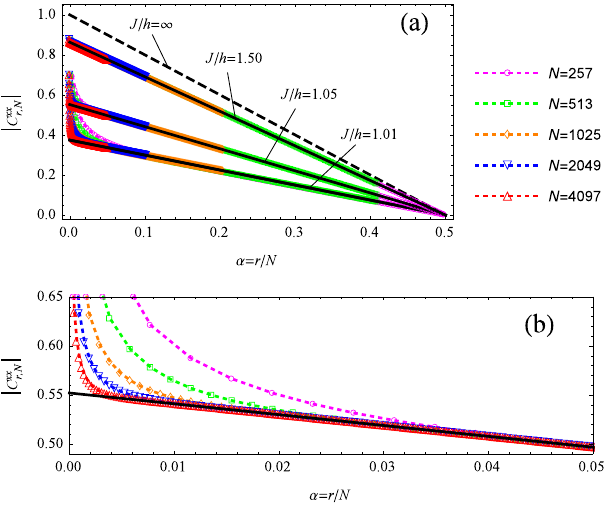}
\end{center}
\caption{Longitudinal correlation functions for several selected values of
$J/h$ in the gapless region. The dashed black line in (a) is an exact result,
(\ref{Cxx(h=0)}), for $J/h=\infty$ (i.e. $h=0$). In both (a) and (b), the
colored dingbat data are direct rigorous evaluations of (\ref{Cxxdet}), while
the black straight lines show the asymptotic behavior described by
(\ref{Cxx2}). (b) is a zoom-in plot for the case $J/h=1.05$.}%
\label{fig7}%
\end{figure}
%%%%%%%%%%% FIG 7  %%%%%%%%%%%%%%%%%%%%%%%%%%%%%%%

First, let us see an exact result in the case $h=0$. By (\ref{Drgapless}), we
have $D_{r}=-\delta_{r,0}$. Then (\ref{Cxxdet1}) is reduced to%
\begin{equation}
C_{r,N}^{xx}=(-1)^{r}(1-2\alpha), \label{Cxx(h=0)}%
\end{equation}
where $\alpha=\frac{r}{N}$. If one takes the limit $N\rightarrow\infty$ first
and gets $C_{r,N}^{xx}\approx(-1)^{r}$, one would think this is a simple
antiferromagnets. But if we take a value of $\alpha\in(0,1/2)$, we see the
exact result, (\ref{Cxx(h=0)}), measures a non-local correlation because
$r=\alpha N\rightarrow\infty$. Please notice that the ground state,
(\ref{sum of kink states}), is an exact superposition of kink states, whose
correlation function is exactly given by (\ref{Cxx(h=0)}). This is purely a
theoretical consequence of the model. The exact result is shown as the dashed
black line in Fig. \ref{fig7}(a).

Second, we work out the asymptotic behaviour for arbitrary $J$ $>h$. For a
large enough system, we can substitute the sum in (\ref{Drgapless}) with an
integral to get%
\begin{equation}
D_{r}\overset{N\rightarrow\infty}{\longrightarrow}\int_{-\pi}^{\pi}\frac
{dq}{2\pi}\exp\left(  -\mathrm{i}qr\right)  D(\mathrm{e}^{\mathrm{i}q}).
\end{equation}
Now we need to evaluate the new type of Toeplitz determinant in Eq.
(\ref{Cxxdet1}). Following the earlier procedure by McCoy and Wu
\cite{Wu,McCoy,McCoy Wu}, we have proved a theorem for this special case
in Appendix A:

\textit{Theorem: Consider a Toeplitz determinant} $\Theta(r,N)$ \textit{in}
(\ref{Cxxdet1}) \textit{with}
\begin{equation}
D_{n}=\int_{-\pi}^{\pi}\frac{dq}{2\pi}\,D(\mathrm{e}^{\mathrm{i}%
q})\,\mathrm{e}^{-\mathrm{i}qn}.
\end{equation}
\textit{If the generating function }$D(\mathrm{e}^{\mathrm{i}q}%
)$\textit{\ and }$\ln D(\mathrm{e}^{\mathrm{i}q})$\textit{\ are
continuous on the unit circle }$\left\vert \mathrm{e}^{\mathrm{i}%
q}\right\vert =1$\textit{, then the behavior for large }$N$\textit{\ and
}$\mathit{r}$\textit{\ of }$\Theta(r,N)$\textit{\ is given by }%
\begin{equation}
\Theta(r,N)=\Delta_{r}(1+\frac{2\alpha}{D(\mathrm{e}^{\mathrm{i}%
0})}),
\end{equation}
\textit{\ where}%
\begin{eqnarray}
\Delta_{r}  & =\mu^{r}\exp(\sum_{n=1}^{\infty}nd_{-n}d_{n}),\\
\mu & =\exp[\int_{-\pi}^{\pi}\frac{dq}{2\pi}\,\ln D(\mathrm{e}%
^{\mathrm{i}q})],\\
d_{n}  & =\int_{-\pi}^{\pi}\frac{dq}{2\pi}\,\mathrm{e}%
^{-\mathrm{i}qn}\ln D(\mathrm{e}^{\mathrm{i}q}),
\end{eqnarray}
\textit{if the sum }$\sum_{n=1}^{\infty}nd_{-n}d_{n}$\textit{ is convergent.}

By applying the above theorem to the gapless region ($J$ $>h$), we get an
asymptotic behavior
\begin{equation}
C_{r,N}^{xx}=(-1)^{r}(1-\frac{h^{2}}{J^{2}})^{1/4}(1-2\alpha). \label{Cxx2}%
\end{equation}
It is clear (\ref{Cxx2}) coincides with (\ref{Cxx(h=0)}). This asymptotic
behavior is depicted in Fig. \ref{fig7}, which is perfectly coincident with
the direct evaluations of (\ref{Cxxdet}). This surprising result is totally
different from the conventional findings \cite{Sachdev}.

\subsection{Entanglement entropy of the ground state}

\begin{figure}[ptb]
\begin{center}
\includegraphics[width=0.7\textwidth]{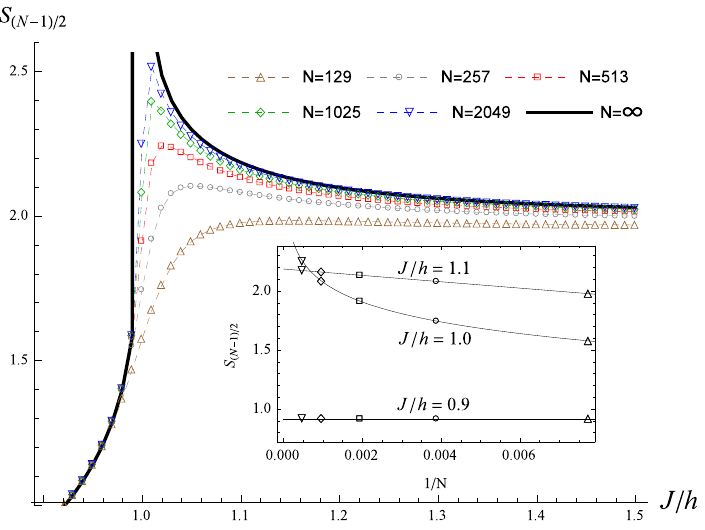}
\end{center}
\caption{(Color
online) Entanglement entropy $S_{(N-1)/2}$ as a function of $J/h$ for a
sequence of number of lattice sites $N$. The inset shows examples of finite
size scaling for extrapolating to $N\rightarrow\infty$. At the critical point,
the numerical data fit a divergent behavior, $S_{(N-1)/2}\sim\frac{1}{6}%
\log_{2}N$, coincident with the prediction by CFT \cite{Holzhey}.}%
\label{fig8}%
\end{figure}

Entanglement entropy is another powerful quantity for exhibiting the entangled
nature of a system. We define the reduced density matrix $\rho_{l}$%
=tr$_{N-l}|E_{0}^{(O,o)}\rangle\langle E_{0}^{(O,o)}|$ and the entanglement
entropy (EE) $S_{l}$=$-$tr$(\rho_{l}\log_{2}\rho_{l})$, where the trace is
performed on the spin states of contiguous sites from $j$=$1$ to $N-l$. We can
evaluate the EE numerically by utilizing the matrix
\cite{Vidal,Latorre,Amico-RMP}
\begin{equation}
\Gamma_{l}=\left\vert
\begin{array}
[c]{cccc}%
\Pi_{0} & \Pi_{1} & \cdots & \Pi_{l-1}\\
\Pi_{-1} & \Pi_{0} & \cdots & \Pi_{l-2}\\
\cdots & \cdots & \cdots & \cdots\\
\Pi_{1-l} & \Pi_{2-l} & \cdots & \Pi_{0}%
\end{array}
\right\vert \mathrm{ with }\Pi_{l}=\left\vert
\begin{array}
[c]{cc}%
0 & -g_{l}\\
g_{-l} & 0
\end{array}
\right\vert ,
\end{equation}
where $g_{l}$=$\mathscr{D}_{l-1}$+$\frac{2}{N}$. Let $V\in SO(2l)$ denote an
orthogonal matrix that brings $\Gamma_{l}$ into a block diagonal form such
that $\Gamma_{l}^{C}$=$V\Gamma_{l}V^{T}$=$\bigoplus_{m=0}^{l-1}($i$v_{m}%
\sigma_{y})$ with $v_{m}\geq0$. Then $S_{l}$ is given by $S_{l}$=$\sum
_{m=0}^{l-1}H_{2}(\frac{1+v_{m}}{2})$ with $H_{2}(x)$=$-x\log_{2}%
x-(1-x)\log_{2}(1-x)$. The numerical results for $l$=$(N-1)/2$ are shown in
Fig. \ref{fig8}. We observe the EE in the gapped region ($J<h$) is small until
near the critical point, where it abruptly tends to become divergent as
predicted by CFT \cite{Holzhey,Korepin,Calabrese,Lin}. While in the gapless
region ($J>h$), we observe $S_{(N-1)/2}$ with $N\rightarrow\infty$ approaches
its minimal value $2$ when $h\rightarrow0$. In fact, one can verify that
$|E_{0}^{(O,o)}\rangle$ evolves with $h\rightarrow0$ adiabatically into a
superposition of all kink states in Eq. (\ref{sum of kink states}), whose EE
is exactly $2$. As a comparison, the EE of the well-known GHZ state is
$\log_{2}2$=$1$.

%%%%%%%%%%% FIG 8  %%%%%%%%%%%%%%%%%%%%%%%%%%%%%%%
%%%%%%%%%%% FIG 8  %%%%%%%%%%%%%%%%%%%%%%%%%%%%%%%

\subsection{Finite temperature properties in the gapless region}

In the gapless region, the dispersed but neatly aligned lowest $2N$ states,
(\ref{state E0})-(\ref{state EOo}), dominate the system's properties at low
temperatures ($T$ $\ll4h/k_{B}$), where $k_{B}$ is the Boltzmann constant.
This fact facilitates us to work out some quantities at low temperatures based
on the partition function
\begin{equation}
Z=\sum_{q\in q^{(O,o)}\cup q^{(O,e)}}\mathrm{e}^{-\beta E_{q}},
\end{equation}
where $\beta=\frac{1}{k_{B}T}$. The DOS is defined as
\begin{equation}
\rho(E)=\frac{1}{N}\sum_{q\in q^{(O,o)}\cup q^{(O,e)}}\,\delta(E-E_{q}).
\end{equation}
If $N\rightarrow\infty$, the summation in the $1$st BZ can be replaced with
integral, so we get a DOS,%
\begin{equation}
\rho(x)=\frac{4(x+2J-2h)}{\pi\sqrt{x(x-4h)(4h-x-4J)(x+4J)}},
\end{equation}
where $x=E-E_{0}$. It can be expanded as%
\begin{equation}
\rho(x)=ax^{-1/2}+bx^{1/2}+O(x^{3/2})
\end{equation}
with
\begin{equation}
a=\frac{(J-h)^{1/2}}{\pi(Jh)^{1/2}},b=\frac{(h^{2}+Jh+J^{2})}{8\pi
(Jh)^{3/2}(J-h)^{1/2}}.
\end{equation}
So we get the specific heat per site at low temperature,
\begin{equation}
\frac{C_{M}(T)}{N}\approx\frac{k_{B}}{2}\left[  1+\frac{2bk_{B}T(4a+bk_{B}%
T)}{(2a+bk_{B}T)^{2}}\right]  . \label{CM}%
\end{equation}

\section{Experimental proposal}

\label{sec:exp}

We can design a large enough one to see the effect of ring frustration with
nowadays state-of-art techniques based on laser-cooled and trapped atomic
ions. In fact, the case for $N=3$ has been experimentally realized
\cite{Edwards}. To generate a system with larger $N\in Odd$ and ensure that
the frustration comes from the ring geometry not from short-range
interactions, we provide another proposal.

\begin{figure}[ptb]
\begin{center}
\includegraphics[width=0.7\textwidth]{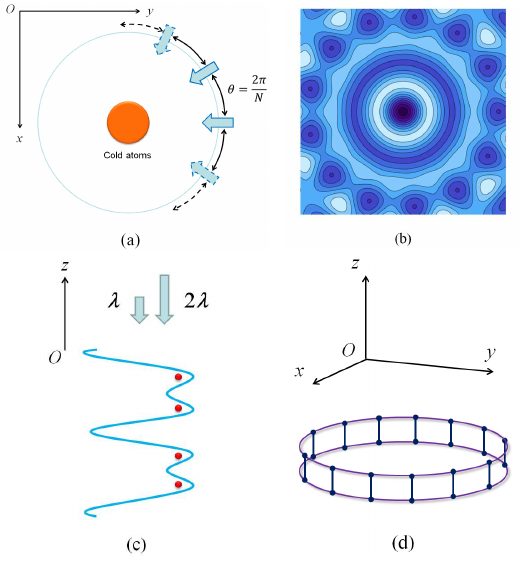}
\end{center}
\caption{(a) Scheme of the proposed experimental setup in $x$-$y$ plane. Each
arrow depicts a wave vector of a standing-wave laser. The angle between any
two neighboring lasers is $2\pi/N$. (b) The exemplified color map of optical
potential where a ring of $13$ trapping wells is shown by the dark blue
potential wells. (c) The arrangement of lasers in $z$ direction, where two
standing wave lasers form an isolated double wells potential. (d) The total
two-leg ladder potential.}%
\label{fig9}%
\end{figure}

In our proposal as shown in Fig. \ref{fig9}, there are two key points. The
first point is to produce a ring potential with odd number of traps. In
$x$-$y$ plane, we impose $N$ beams of independent standing wave lasers which
are obtained by frequency selection. Then, each standing wave will contributes
an optical potential along $\overrightarrow{k_{i}}$\ direction that can be
expressed as $V_{x\mathrm{-}y}\cos^{2}(\overrightarrow{k_{i}}\cdot
\overrightarrow{r_{i}}-\phi)$\ for the i-th beam, where $\overrightarrow{k_{i}%
}$ is the strength of beams and $\phi$ is the phase shift. The angle between
two neighboring lasers is $2\pi/N$. Thus, by adopting appropriate
$V_{x\mathrm{-}y}$\ and $\phi$, we can obtain a circular lattice potential with
$N$ traps in $x$-$y$ plane (Fig. \ref{fig9}(a) and (b)).

The second point is to realize the antiferromagnetic transverse model
robustly. In $z$ direction we apply two independent standing wave lasers,
$V_{z1}\cos^{2}(k_{z}z)$ and $V_{z2}\cos^{2}(2k_{z}z)$, where the former has
twice wave length of the latter. Eventually, we obtain a periodical two-leg
ladder potential by forming a double-well potential in $z$ direction (Fig.
\ref{fig9}(c) and (d)). In real experiment, there is additional harmonic
trapping potential $V_{trap}\left(  x^{2}+y^{2}\right)  $. The total potential
can be written as%
\begin{eqnarray}
V\left(  x,y,z\right)   &  =V_{trap}\left(  x^{2}+y^{2}\right)  +V_{z1}%
\cos^{2}\left(  k_{z}z\right)  +V_{z2}\cos^{2}\left(  2k_{z}z\right)
\nonumber\\
&  +V_{x\mathrm{-}y}\sum_{i=1}^{N}\cos^{2}\left(  \overrightarrow{k_{i}}%
\cdot\overrightarrow{r_{i}}-\phi\right)  .
\end{eqnarray}

%%%%%%%%%%% FIG 9  %%%%%%%%%%%%%%%%%%%%%%%%%%%%%%%
%%%%%%%%%%% FIG 9  %%%%%%%%%%%%%%%%%%%%%%%%%%%%%%%

Then, let us consider loading into the ladders with cold atoms
which have two relevant internal states denoted as pseudo-spin states
$\lambda=\uparrow,\downarrow$. The lattice potential experienced by cold atoms
depends on which of those two internal states are located. For sufficiently
deep potential and low temperatures, the system will be described by the
following bosonic or fermionic Hubbard model \cite{Duan},%
\begin{eqnarray}
H_{Hub}  &  =\sum_{j,\lambda,s}(-t_{\lambda})(a_{j\lambda,s}^{\dag
}a_{(j+1)\lambda,s}+\mathrm{h.c.})+\sum_{j,\lambda}(-t_{\lambda})(a_{j\lambda
,1}^{\dag}a_{j\lambda,2}+\mathrm{h.c.})\nonumber\\
&  +\frac{1}{2}\sum_{j,\lambda,s}U_{\lambda}n_{j\lambda,s}(n_{j\lambda
,s}-1)+\sum_{j,s}U_{\uparrow\downarrow}n_{j\uparrow,s}n_{j\downarrow,s},
\label{exphhub}%
\end{eqnarray}
where $s=1,2$ is the leg index. With the conditions of Mott insulator limit
$t_{\lambda}\ll U_{\lambda}$, $U_{\uparrow\downarrow}$ and half filling
$\langle n_{j\uparrow,s}\rangle+\langle n_{j\downarrow,s}\rangle\approx1$, the
low-energy Hamiltonian of (\ref{exphhub}) is mapped to the $\mathrm{XXZ}$ model
by second-order perturbation,%
\begin{eqnarray}
H_{s}  &  =\sum_{j,s}\pm J_{\perp}(S_{j,s}^{x}S_{j+1,s}^{x}+S_{j,s}%
^{y}S_{j+1,s}^{y})+J_{z}S_{j,s}^{z}S_{j+1,s}^{z}\nonumber\\
&  +\sum_{j}\pm K_{\perp}(S_{j,1}^{x}S_{j,2}^{x}+S_{j,1}^{y}S_{j,2}^{y}%
)+K_{z}S_{j,1}^{z}S_{j,2}^{z}, \label{exphspin}%
\end{eqnarray}
where the pseudo-spin operator $\mathbf{S}=a^{\dag}\overrightarrow{\sigma}%
a/2$, $\overrightarrow{\sigma}=(\sigma_{x},\sigma_{y},\sigma_{z})$ are the
Pauli matrices and $a^{\dag}=\left(  a_{\uparrow}^{\dag},a_{\downarrow}^{\dag
}\right)  $. The positive signs before $J_{\perp},K_{\perp}$ are for fermionic
atoms and negative signs for bosonic one. The interaction coefficients for
bosons are given by,
\begin{eqnarray*}
J_{\perp}  &  =\frac{4t_{\uparrow}t_{\downarrow}}{U_{\uparrow\downarrow}%
},J_{z}=\frac{2\left(  t_{\uparrow}^{2}+t_{\downarrow}^{2}\right)
}{U_{\uparrow\downarrow}}-\frac{t_{\uparrow}^{2}}{U_{\uparrow}}-\frac
{t_{\downarrow}^{2}}{U_{\downarrow}},\\
K_{\perp}  &  =\frac{4t_{\uparrow}^{\prime}t_{\downarrow}^{\prime}%
}{U_{\uparrow\downarrow}},K_{z}=\frac{2t_{\uparrow}^{\prime2}+t_{\downarrow
}^{\prime2}}{U_{\uparrow\downarrow}}-\frac{t_{\uparrow}^{\prime2}}%
{U_{\uparrow}}-\frac{t_{\downarrow}^{\prime2}}{U_{\downarrow}}.
\end{eqnarray*}
For fermions, we only need to omit the last two terms in $J_{z}$ and $K_{z}$.
By modulating the intensity, the phase shift of the trapping laser beams, and
the $s$ wave scattering length through Feshbach resonance, we can obtain a
desired Hamiltonian from (\ref{exphspin}),%
\begin{equation}
H_{s}=\sum_{j,s}J_{z}S_{j,s}^{z}S_{j+1,s}^{z}+\sum_{j}K_{{}}\vec{S}_{j,1}%
\cdot\vec{S}_{j,2}.
\end{equation}
The properties of this system are dominated by the pseudo-spin singlet
$\left\vert s\right\rangle _{j}=\left(  \left\vert \uparrow\downarrow
\right\rangle _{j}-\left\vert \downarrow\uparrow\right\rangle _{j}\right)
/\sqrt{2}$ and triplet $\left\vert t_{0}\right\rangle _{j}=\left(  \left\vert
\uparrow\downarrow\right\rangle _{j}+\left\vert \downarrow\uparrow
\right\rangle _{j}\right)  /\sqrt{2}$\ on the rung of the ladders in low
energy. At this time, the system can be mapped to the transverse Ising ring,
(\ref{H1}), that we desired \cite{ChenQH}.

\section{Conclusion and discussion}

\label{sec:conclusion}

In this work we have treated a special system, the transverse Ising ring, with
perfect PBC. The main focus is placed on the case with ring frustration due to
antiferromagnetic seam. We have demonstrated how the fermionic a-cycle problem
is applied for solving the transverse Ising ring based on a method of band
structure analysis. We have shown it is crucial to project out the redundant
DOF of the fermions to restore the full DOF of the original spin model. The
odevity of the number of lattice sites triggers or shuts the presence of ring
frustration. The most intriguing result is that the system in strong
antiferromagnetic region develops a gapless spectrum when the ring frustration
is turned on no matter how large the system is. To the best of our knowledge,
this gapless spectrum is totally unaware in previous investigations. The
non-local nature of the longitudinal correlation function of the ground state
is uncovered in detail. To understand all the fascinating properties of the
system, we presented a treatment of perturbative theory for a simple but
reliable cartoon picture of the formation of gapless spectrum. As an
approximation for large enough system and low enough temperature, the DOS and
specific heat are worked out. We also proposed an experimental protocol for
observing the fascinating phenomenon due to the ring frustration.

There are some issues need to be specified. First, the method of band
structure analysis encounters a problem for evaluating thermodynamic
quantities at arbitrary temperature, because the fermionic a-cycle problem is
not a free fermion one, to which the Fermi distribution can not be applied and
the summation on the constraint fermionic states can not be accomplished in a
closed form. But for finite system, it is just a matter of amount of
computation. The finite system can be utilized to produce highly entangled
states. Second, the method for the a-cycle problem can not be applied to the
system with OBC, where the c-cycle problem is applicable.

The odevity-induced phenomenon is reminiscent of the one in the well-known
spin ladders \cite{Dagotto}. But the situation here is more dramatical because
the difference between the consequences of $N\in Even$ and $N\in Odd$ is
robust even when $N\rightarrow\infty$.

We acknowledge useful discussions with Yan He. This work was supported by the
NSFC under Grants no. 11074177, SRF for ROCS SEM (20111139-10-2).

\appendix

\section{Proof of a Theorem}

\label{sec:cor}

\noindent\textit{Theorem: Consider a Toeplitz determinant}%
\begin{equation}
\Theta(r,N)=\left\vert
\begin{array}
[c]{cccc}%
D_{0}+\frac{2}{N} & D_{-1}+\frac{2}{N} & \cdots & D_{-r+1}+\frac{2}{N}\\
D_{1}+\frac{2}{N} & D_{0}+\frac{2}{N} & \cdots & D_{-r+2}+\frac{2}{N}\\
\cdots & \cdots & \cdots & \cdots\\
D_{r-1}+\frac{2}{N} & D_{r-2}+\frac{2}{N} & \cdots & D_{0}+\frac{2}{N}%
\end{array}
\right\vert \label{sTheta}%
\end{equation}
\textit{with }$D_{n}$\textit{=}$\int_{-\pi}^{\pi}\frac{dq}{2\pi}%
\,D(\mathrm{e}^{\mathrm{i}q})\,\mathrm{e}%
^{-\mathrm{i}qn}$. \textit{If the generating function }%
$D(\mathrm{e}^{\mathrm{i}q})$\textit{ and }$\ln
D(\mathrm{e}^{\mathrm{i}q})$\textit{\ are continuous on the unit
circle }$\left\vert \mathrm{e}^{\mathrm{i}q}\right\vert
=1$\textit{, then the behavior for large }$N$\textit{ and }$\mathit{r}%
$\textit{ of }$\Theta(r,N)$\textit{ is given by}%
\begin{equation}
\Theta(r,N)=\Delta_{r}(1+\frac{2\alpha}{D(\mathrm{e}^{\mathrm{i}%
0})}),
\end{equation}
\textit{\ where }$\alpha=\frac{r}{N}$, $\Delta_{r}=\mu^{r}\exp(\sum
_{n=1}^{\infty}nd_{-n}d_{n})$, $\mu=\exp[\int_{-\pi}^{\pi}\frac{dq}{2\pi}\,\ln
D(\mathrm{e}^{\mathrm{i}q})]$, and $d_{n}=\int_{-\pi}^{\pi}%
\frac{dq}{2\pi}\,\mathrm{e}^{-\mathrm{i}qn}\ln D(\mathrm{e}%
^{\mathrm{i}q})$, \textit{if the sum }$\sum_{n=1}^{\infty}nd_{-n}d_{n}%
$\textit{ is convergent.}

\strut\noindent\textit{Proof: }Let\textit{\ }$\mathrm{e}%
^{\mathrm{i}q}=\xi,$ $D_{n}=\int_{-\pi}^{\pi}\frac{dq}{2\pi}\,D(\xi
)\xi^{-n}\,$. First, we rewrite Eq. (\ref{sTheta}) as%
\begin{eqnarray}
\fl \Theta(r,N)   =\left\vert
\begin{array}
[c]{cccc}%
D_{0} & D_{-1} & \cdots & D_{-r+1}\\
D_{1} & D_{0} & \cdots & D_{-r+2}\\
\cdots & \cdots & \cdots & \cdots\\
D_{r-1} & D_{r-2} & \cdots & D_{0}%
\end{array}
\right\vert +\left\vert
\begin{array}
[c]{cccc}%
\frac{2}{N} & D_{-1} & \cdots & D_{1-r}\\
\frac{2}{N} & D_{0} & \cdots & D_{2-r}\\
\cdots & \cdots & \cdots & \cdots\\
\frac{2}{N} & D_{r-2} & \cdots & D_{0}%
\end{array}
\right\vert \nonumber\\
  +\ldots+\left\vert
\begin{array}
[c]{cccc}%
D_{0} & \frac{2}{N} & \cdots & D_{2-r}\\
D_{1} & \frac{2}{N} & \cdots & D_{2-r}\\
\cdots & \cdots & \cdots & \cdots\\
D_{r-1} & \frac{2}{N} & \cdots & D_{0}%
\end{array}
\right\vert +\left\vert
\begin{array}
[c]{cccc}%
D_{0} & D_{-1} & \cdots & \frac{2}{N}\\
D_{1} & D_{0} & \cdots & \frac{2}{N}\\
\cdots & \cdots & \cdots & \cdots\\
D_{r-1} & D_{r-2} & \cdots & \frac{2}{N}%
\end{array}
\right\vert
\end{eqnarray}
Then we compose a set of linear equations
\begin{equation}
\sum_{m=0}^{r-1}D_{n-m}x_{m}^{(r-1)}=\frac{2}{N}\;,\;0\leq n\leq
r-1.\label{sumeq}%
\end{equation}
These equations have an unique solution for $x_{n}^{(r-1)}$ if there exists a
non-zero determinant:
\begin{equation}
\Delta_{r}\equiv\left\vert
\begin{array}
[c]{cccc}%
D_{0} & D_{-1} & \cdots & D_{1-r}\\
D_{1} & D_{0} & \cdots & D_{2-r}\\
\cdots & \cdots & \cdots & \cdots\\
D_{r-1} & D_{r-2} & \cdots & D_{0}%
\end{array}
\right\vert \neq0.
\end{equation}
By Cramer's rule, we have the solution:
\begin{eqnarray}
x_{0}^{(r-1)}   =\frac{\left\vert
\begin{array}
[c]{cccc}%
\frac{2}{N} & D_{-1} & \cdots & D_{1-r}\\
\frac{2}{N} & D_{0} & \cdots & D_{2-r}\\
\cdots & \cdots & \cdots & \cdots\\
\frac{2}{N} & D_{r-2} & \cdots & D_{0}%
\end{array}
\right\vert }{\Delta_{r}}\;\\
x_{1}^{(r-1)}   =\frac{\left\vert
\begin{array}
[c]{cccc}%
D_{0} & \frac{2}{N} & \cdots & D_{2-r}\\
D_{1} & \frac{2}{N} & \cdots & D_{2-r}\\
\cdots & \cdots & \cdots & \cdots\\
D_{r-1} & \frac{2}{N} & \cdots & D_{0}%
\end{array}
\right\vert }{\Delta_{r}}\;\\
 \quad\quad\quad\quad\quad\quad\quad\quad\quad\quad\vdots\\ \nonumber
x_{r-1}^{(r-1)}   =\frac{\left\vert
\begin{array}
[c]{cccc}%
D_{0} & D_{-1} & \cdots & \frac{2}{N}\\
D_{1} & D_{0} & \cdots & \frac{2}{N}\\
\cdots & \cdots & \cdots & \cdots\\
D_{r-1} & D_{r-2} & \cdots & \frac{2}{N}%
\end{array}
\right\vert }{\Delta_{r}}.
\end{eqnarray}
So we arrive at
\begin{equation}
\Theta(r,N)=\Delta_{r}+\Delta_{r}\sum_{n=0}^{r-1}x_{n}^{(r-1)}.\label{sumxn}%
\end{equation}
For our problem, $\Delta_{r}$ can be evaluated directly by using Szeg\"{o}'s
Theorem, so we need to know how to calculate the second term in Eq.
(\ref{sumxn}). Follow the standard Wiener-Hopf procedure \cite{McCoy Wu,McCoy,
Wu}, we consider a generalization of Eq. (\ref{sumeq})
\begin{equation}
\sum_{m=0}^{r-1}D_{n-m}x_{m}=y_{n}\mathrm{, \ }0\leq n\leq r-1\label{Wiener}%
\end{equation}
and define
\begin{equation}
x_{n}=y_{n}=0\quad\mathrm{for}\quad n\leq-1\quad\mathrm{and}\quad n\geq r
\end{equation}%
\begin{eqnarray}
v_{n}   =\sum_{m=0}^{r-1}D_{-n-m}x_{m}\quad\mathrm{for}\quad n\geq1\nonumber\\
 =0\quad\mathrm{for}\quad n\leq0
\end{eqnarray}%
\begin{eqnarray}
u_{n}   =\sum_{m=0}^{r-1}D_{r-1+n-m}x_{m}\quad\mathrm{for}\quad n\geq
1\nonumber\\
  =0\quad\mathrm{for}\quad n\leq0
\end{eqnarray}
We further define
\begin{eqnarray}
D\left(  \xi\right)     =\sum_{n=-\infty}^{\infty}D_{n}\xi^{n},\quad Y\left(
\xi\right)  =\sum_{n=0}^{r-1}y_{n}\xi^{n},\quad V\left(  \xi\right)
=\sum_{n=1}^{\infty}v_{n}\xi^{n},\quad\nonumber\\
U\left(  \xi\right)     =\sum_{n=1}^{\infty}u_{n}\xi^{n},\quad X\left(
\xi\right)  =\sum_{n=0}^{r-1}x_{n}\xi^{n}.\label{fourier}%
\end{eqnarray}
It then follows from Eq. (\ref{Wiener}) that we can get
\begin{equation}
D\left(  \xi\right)  X\left(  \xi\right)  =Y\left(  \xi\right)  +V\left(
\xi^{-1}\right)  +U\left(  \xi\right)  \xi^{r-1}\label{fourier form}%
\end{equation}
for $|\xi|=1$. Becuase $D\left(  \xi\right)  $ and $\ln D\left(  \xi\right)  $
is continuous and periodic on the unit circle, $D\left(  \xi\right)  $ has a
unique factorization, up to a multiplicative constant, in the form
\begin{equation}
D\left(  \xi\right)  =P^{-1}\left(  \xi\right)  Q^{-1}\left(  \xi^{-1}\right)
,\label{Dxi}%
\end{equation}
for $|\xi|=1$, such that $P\left(  \xi\right)  $ and $Q\left(  \xi\right)  $
are both analytic for $|\xi|<1$ and continuous and nonzero for $|\xi|\leq1$.
we may now use the factorization of $D\left(  \xi\right)  $ in Eq.
(\ref{fourier form}) to write
\begin{eqnarray}
\fl  P^{-1}\left(  \xi\right)  X\left(  \xi\right)  -\left[  Q\left(  \xi
^{-1}\right)  Y\left(  \xi\right)  \right]  _{+}-\left[  Q\left(  \xi
^{-1}\right)  U\left(  \xi\right)  \xi^{r-1}\right]  _{+}\nonumber\\
 \fl \quad =\left[  Q\left(  \xi^{-1}\right)  Y\left(  \xi\right)  \right]
_{-}+Q\left(  \xi^{-1}\right)  V\left(  \xi^{-1}\right)  +\left[  Q\left(
\xi^{-1}\right)  U\left(  \xi\right)  \xi^{r-1}\right]  _{-},\label{pq}%
\end{eqnarray}
where the subscript $+\left(  -\right)  $ means that we should expand the
quantity in the brackets into a Laurent series and keep only those terms where
$\xi$ is raised to a non-negative (negative) power. The left-hand side of Eq.
(\ref{pq}) defines a function analytic for $|\xi|<1$ and continuous on
$|\xi|=1$ and the right-hand side defines a function which is analytic for
$|\xi|>1$ and is continuous for $|\xi|=1$. Taken together they define a
function $E(\xi)$ analytic for all $\xi$ except possibly for $|\xi|=1$ and
continuous everywhere. But these properties are sufficient to prove that
$E(\xi)$ is an entire function which vanished at $|\xi|=\infty$ and thus, by
Liouville's theorem, must be zero everywhere \cite{McCoy Wu,McCoy}. Therefore
both the right-hand side and the left-hand side of Eq. (\ref{pq}) vanish
separately and thus we have
\begin{equation}
X\left(  \xi\right)  =P\left(  \xi\right)  \left\{  \left[  Q\left(  \xi
^{-1}\right)  Y\left(  \xi\right)  \right]  _{+}+\left[  Q\left(  \xi
^{-1}\right)  U\left(  \xi\right)  \xi^{r-1}\right]  _{+}\right\}  .
\end{equation}
Furthermore, $U\left(  \xi\right)  $ can be neglected for large r%
\begin{equation}
X\left(  \xi\right)  \approx P\left(  \xi\right)  \left[  Q\left(  \xi
^{-1}\right)  Y\left(  \xi\right)  \right]  _{+}.\label{xxi}%
\end{equation}
Consider the term $\left[  Q\left(  \xi^{-1}\right)  Y\left(  \xi\right)
\right]  _{+}$, because $Q\left(  \xi\right)  $ is a $+$ function, so we can
expand it as a Laurent series and keep only those term where $\xi$ is raised
to a non-negative power,
\begin{equation}
\fl Q\left(  \xi\right)  ={\sum\limits_{n=0}^{\infty}}a_{n}\xi^{n}=\left(
a_{0}+a_{1}\xi^{1}+a_{2}\xi^{2}+\cdots+a_{r-1}\xi^{r-1}\right)  +O\left(
\xi^{r}\right)  ,
\end{equation}
and then
\begin{equation}
Q\left(  \xi^{-1}\right)  =a_{0}+a_{1}\xi^{-1}+a_{2}\xi^{-2}+\cdots+a_{r-1}%
\xi^{1-r},
\end{equation}
where we have neglected the term $O\left(  \xi^{r}\right)  $\ for large $r$
for clarity. From Eq. (\ref{sumeq}) and Eq. (\ref{fourier}), we have
\begin{equation}
Y\left(  \xi\right)  =\sum_{n=0}^{r-1}y_{n}\xi^{n}=\frac{2}{N}\left(
1+\xi^{1}+\xi^{2}+\cdots+\xi^{r-1}\right)  ,
\end{equation}
thus
\begin{eqnarray}
\fl \left[  Q\left(  \xi^{-1}\right)  Y\left(  \xi\right)  \right]
_{+}\nonumber\\
\fl \quad =\frac{2}{N}[\left(  a_{0}+a_{1}+a_{2}+\cdots+a_{r-1}\right)  +\left(
a_{0}+a_{1}+\cdots+a_{r-2}\right)  \xi^{1}+\cdots+a_{0}\xi^{r-1}].
\end{eqnarray}
From Eq. (\ref{sumxn}), Eq. (\ref{fourier}) and Eq. (\ref{xxi}), we have%
\begin{equation}
\sum_{n=0}^{r-1}x_{n}^{(r-1)}=X\left(  1\right)  =P\left(  1\right)  \left[
Q\left(  1\right)  Y\left(  1\right)  \right]  _{+},
\end{equation}%
\begin{eqnarray}
\left[  Q\left(  1\right)  Y\left(  1\right)  \right]  _{+}  =\frac{2}%
{N}\left[  r\left(  a_{0}+a_{1}+a_{2}+\cdots+a_{r-1}\right)  \right]
\nonumber\\
  -\frac{2}{N}\left[  a_{1}+2a_{2}+\cdots+\left(  r-1\right)  a_{r-1}\right]
\\
  =\frac{2}{N}\left[  rQ\left(  1\right)  -\frac{dQ\left(  \xi\right)  }%
{d\xi}|_{\xi=1}\right]  .\label{qy}%
\end{eqnarray}
So when $r\gg1$, we can ignore the second term in Eq. (\ref{qy}). Together
with Eq. (\ref{Dxi}), we get
\begin{equation}
X\left(  1\right)  =\frac{2r}{N}P\left(  1\right)  Q\left(  1\right)
=\frac{2r}{ND\left(  1\right)  }=\frac{2r}{ND\left(  \mathrm{e}%
^{\mathrm{i}0}\right)  }.
\end{equation}
At last, by Szeg\"{o}'s Theorem, we get
\begin{equation}
\Delta_{r}=\mu^{r}\exp(\sum_{n=1}^{\infty}nd_{-n}d_{n}),
\end{equation}
where%
\begin{equation}
\mu=\exp\left[  \int_{-\pi}^{\pi}\frac{dq}{2\pi}\,\ln D(\mathrm{e}%
^{\mathrm{i}q})\right]  ,d_{n}=\int_{-\pi}^{\pi}\frac{dq}{2\pi
}\,e^{-\mathrm{i}qn}\ln D(\mathrm{e}^{\mathrm{i}%
q}).\nonumber
\end{equation}
From Eq. (\ref{sumxn}), we have%
\begin{equation}
\Theta(r,N)=\Delta_{r}(1+\frac{2\alpha}{D(\mathrm{e}^{\mathrm{i}%
0})}),\alpha=\frac{r}{N}.
\end{equation}
Q.E.D.

\end{document}